\documentclass[11pt]{article}
\usepackage{xspace}
\usepackage{graphicx}
\usepackage{amsmath}
\usepackage{amssymb}
\usepackage{color}
\usepackage{caption}
\usepackage{subfigure}
\usepackage{gensymb}

\definecolor{Red}{rgb}{1,0,0}
\definecolor{Green}{rgb}{0,1,0}
\definecolor{Blue}{rgb}{0,0,1}
\definecolor{Black}{rgb}{0,0,0}

\def\beq{\begin{equation}}
\def\eeq#1{\label{#1}\end{equation}}
\def\eeqn{\end{equation}}

\def\beqa{\begin{eqnarray}}
\def\eeqa#1{\label{#1}\end{eqnarray}}
\def\eeqan{\end{eqnarray}}

\let\bar=\overbar

\def\Dslash{\not{\hbox{\kern-4pt $D$}}}
\def\dslash{\not{\hbox{\kern-2pt $\del$}}}

\def\msb{{\bar{\ssstyle M \kern -1pt S}}}

\def\Title#1{\begin{center} {\Large {\bf #1} } \end{center}}

\begin{document}

\Title{Investigating the differences between electron and muon neutrino interactions using the T2K near detector.}

\bigskip\bigskip


\begin{raggedright}  

{\it Iain Lamont\index{Lamont, I.},\\
Department of Physics\\
Lancaster University\\
LA1 4YW Lancaster, UK}\\

\end{raggedright}

\section{Abstract}
The T2K neutrino beam consists mostly of muon neutrinos with a 1$\%$ component of electron neutrinos \cite{beam}. In order to maximise the physics potential of T2K and other future neutrino experiments, it is important to understand how these electron and muon neutrinos interact. To this end, the ratio of the Charged-Current Quasi-Elastic (CCQE) cross section to the total Charged Current (CC) cross section is taken for both $\nu_{e}$ and $\nu_\mu$ using data from the T2K near detector, ND280, and simulated data from NEUT \cite{Hayato:2002sd} and GENIE \cite{Andreopoulos:2009zz} Monte Carlo generators. This has the advantage that many of the systematic uncertainties will cancel in the analysis, including the flux of $\nu_{e}$ and $\nu_\mu$ in the beam. The double ratio of these two ratios is then taken as a means of directly comparing the interactions of the two neutrino flavours.

\section{The T2K Experiment}

The T2K experiment is a long-baseline neutrino experiment that uses the 30 GeV proton beam at the J-PARC facility in Tokai, Japan to create a beam primarily composed of muon neutrinos. The beam is aligned so that the far detector, Super-Kamiokande, and the off-axis near detector, ND280, are both located 2.5 degrees off-axis, at distances of 295 km and 280 m respectively. By measuring the number of $\nu_{e}$ and $\nu_\mu$ events at the near detector and extrapolating a predicted number of events at the far detector, T2K can study the probability of a $\nu_\mu$ oscillating to a $\nu_{e}$.

\begin{figure}[!ht]
\centering
\includegraphics[width=.4\textwidth]{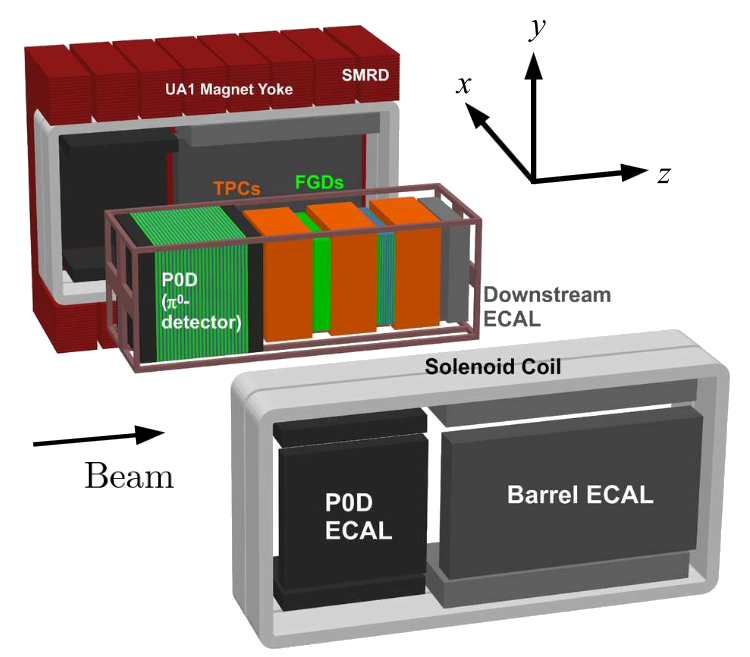}
\caption{An exploded schematic of the T2K near detector, ND280. }
\label{nd280}
\end{figure}

\noindent The ND280 detector has several physics goals. Its primary purpose is to measure the properties and spectrum of the neutrino beam, but it can also be used for independent physics studies such as measurements of the neutrino interaction cross sections. To achieve these goals, the ND280 uses several different subdetectors. Two Fine Grained Detectors (FGDs) are used as target mass for neutrino interactions and for the tracking of particles leaving the interaction. The two FGDs are complemented by three Time Projection Chambers (TPCs) that alternate with the FGDs. These have greater tracking capabilities than the FGDs and combined with a 0.2 T magnetic field allow for momentum and charge determination of charged tracks leaving the FGDs. A dedicated $\pi^{0}$ detector (P0D), located upstream of the FGDs and TPCs, is designed to help understand one of the principal backgrounds at Super-Kamiokande ($\pi^{0} \rightarrow \gamma \gamma$ mimicking $\nu_{e}$ interactions.) Complementing the goals of each of these subdetectors is a suite of Electromagnetic Calorimeters (ECals), which surround the FGDs, TPCs and P0D. These assist in particle identification (PID) and are useful for catching photons as they escape the inner detectors \cite{T2K}. A schematic of the ND280 detector can be seen in Fig.~\ref{nd280}.

\section{Selecting $\nu_{e}$ and $\nu_\mu$ Charged Current Events in the ND280 Tracker}
\label{sec:selections}

\begin{figure}[ht]
\begin{minipage}[b]{0.49\linewidth}
\centering
\includegraphics[width=0.8\textwidth]{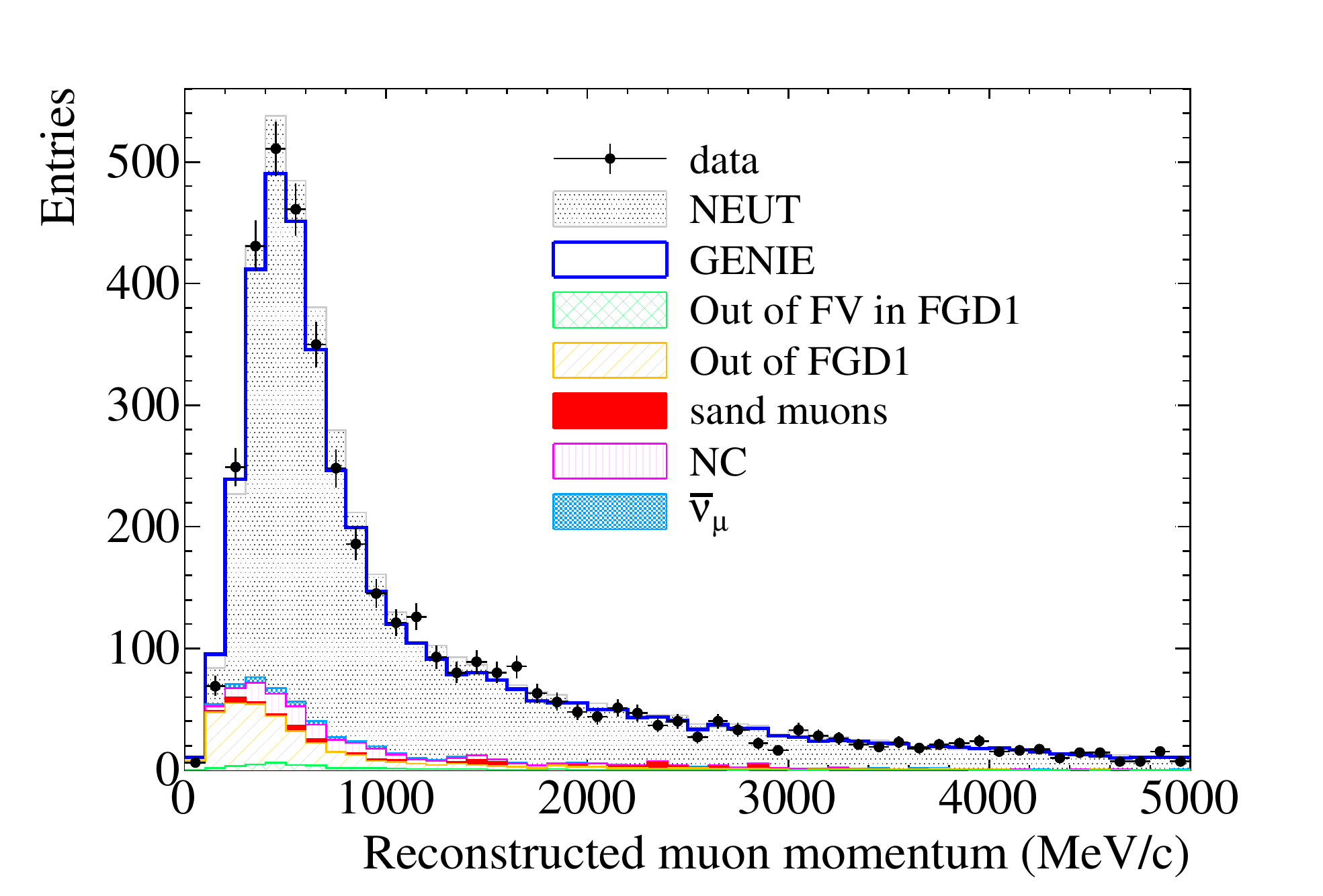}
\hspace{0.5cm}
\end{minipage}
\begin{minipage}[b]{0.49\linewidth}
\centering
\includegraphics[width=1\textwidth]{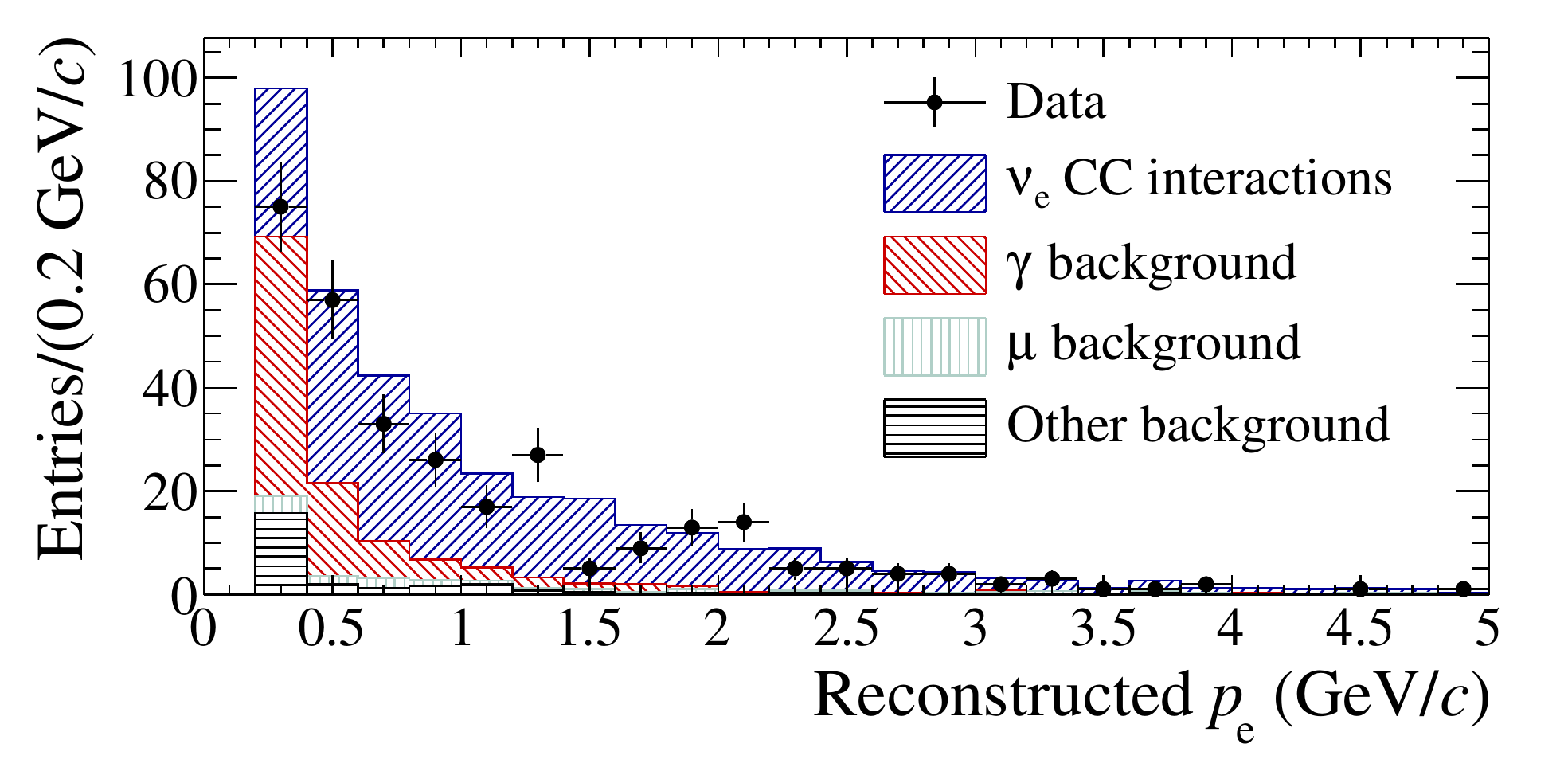}
\end{minipage}
\caption{Reconstructed outgoing lepton momentum distributions for candidate $\nu_\mu$ (left) and $\nu_{e}$ (right) CC interactions in the ND280 detector \cite{Abe:2013jth} \cite{Abe:2014agb}. The points are T2K data, the coloured histograms are the MC predictions.}
\label{selections}
\end{figure}

The selections for $\nu_{e}$ and $\nu_\mu$ charged-current events require a track starting in an FGD and passing through a TPC with enough TPC hits to be classified as good quality. PID criteria are then applied to the track, based on the rate of energy loss of the track (dE/dx), with extra criteria on a track if it enters an ECal. As a means of removing backgrounds, an event is rejected if there is any activity in a TPC, FGD, ECal or the P0D upstream of the interaction FGD. For the enhanced samples to select CCQE events, an additional requirement is applied that the selected track is the only reconstructed track coming from the interaction vertex. Additional cuts are applied to both samples to further reduce the various backgrounds in both samples. The data samples selected using these criteria are shown in Fig.~\ref{selections}. The $\nu_{e}$ selection purity is lower due to a large background from misidentified photon conversions.

\section{Results}
\label{results}

Using the event selections described in Sec.~\ref{sec:selections} for NEUT Monte Carlo fake data, the event ratios for CCQE to CC interactions can be calculated in terms of true neutrino energy. In the future, these quantities can be obtained using data with an unfolding procedure to access neutrino energy. As would be expected, these ratios decrease with energy. This corresponds to CCQE interactions becoming less dominant at higher energies. The advantage of taking these ratios is that the beam flux, number of target nucleons and many of the systematic uncertainties will partially cancel.

\begin{figure}[ht]
\begin{minipage}[b]{0.49\linewidth}
\centering
\includegraphics[width=0.8\textwidth]{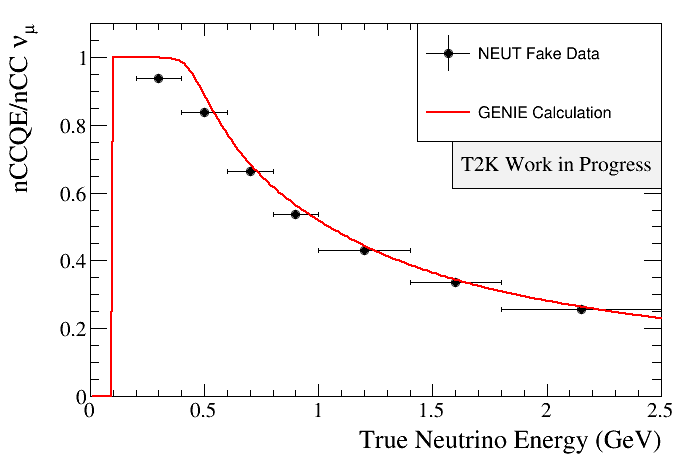}
\hspace{0.5cm}
\end{minipage}
\begin{minipage}[b]{0.49\linewidth}
\centering
\includegraphics[width=0.8\textwidth]{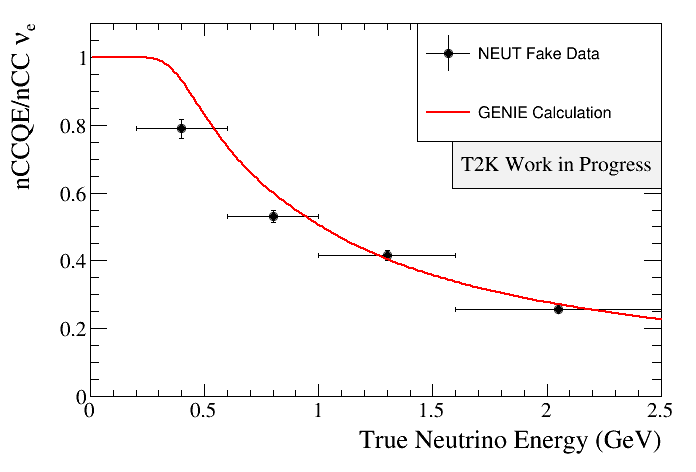}
\end{minipage}
\caption{The ratio of the number of charged-current quasi-elastic events to the total number of charged-current events for $\nu_\mu$ on the left and $\nu_{e}$ on the right. The black points are NEUT fake data corresponding to $4.5 \times 10^{21}$ protons on target (uncertainties are statistical only) whilst the red lines are the calculated ratios using GENIE Monte Carlo.}
\label{single_ratios}
\end{figure}

\begin{figure}[!ht]
\centering
\includegraphics[width=0.6\textwidth]{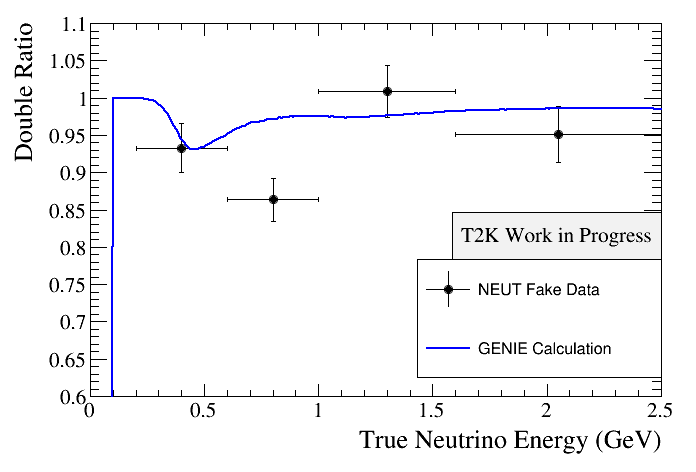}
\caption{The double ratio of CCQE to CC neutrino interaction cross sections, for ($\nu_{e} / \nu_\mu$) (uncertainties are statistical only). Refer to \cite{Day:2012gb} for a description of the differences in $\nu_{e}$ and $\nu_\mu$ CCQE interactions. The dip at $\sim$0.4 GeV in the GENIE calculation corresponds to the higher threshold for a $\nu_\mu$ CC interaction; the expected ratio above this dip remains $<1$ because $\nu_\mu$ CC interactions are relatively more likely to be quasi-elastic.}
\label{double_ratio}
\end{figure}

\noindent The ratio of the two plots in Fig.~\ref{single_ratios} can be used as a direct comparison between electron neutrinos and muon neutrinos, as shown in Fig.~\ref{double_ratio}.

\section{Summary}
This paper presents an initial analysis for a measurement of the difference in the charged-current interactions between $\nu_{e}$ and $\nu_\mu$, based on Monte Carlo simulation. At this point, all errors are statistical only; systematic studies are in progress and will be implemented in the final analysis. In the future, an unfolding procedure will be used on T2K data to convert measured lepton momentum to neutrino energy. Then the event ratios can be calculated for data. This work is currently in progress.

\bigskip
\section{Acknowledgments}

This work was presented on behalf of the T2K collaboration. The presenter would like to thank Dr. Jaroslaw Nowak and the T2K publication board for their input.

\end{document}